\documentclass[aps,prd,twocolumn,groupedaddress,showpacs]{revtex4}

\usepackage[dvips]{graphicx}
\usepackage{amsmath, amssymb}

\newcommand{\mathsym}[1]{{}}

\begin{document}

\title{Stochastic Gravitational Wave Background from Light Cosmic Strings}

\author{Matthew R. DePies}
\email{depies@phys.washington.edu}

\author{Craig J. Hogan}
\email{hogan@u.washington.edu}
\affiliation{University of Washington, Departments of Physics and Astronomy, Seattle WA 98195}

\date{\today}

\begin{abstract}
Spectra of the stochastic gravitational wave backgrounds from cosmic strings are calculated and compared with present and future experimental limits.  Motivated by theoretical expectations of light cosmic strings in superstring cosmology, improvements in experimental sensitivity, and recent demonstrations of large, stable loop formation from a primordial network, this study explores a new range of string parameters with masses lighter than previously investigated.  A standard ``one-scale'' model for string loop formation is assumed.  Background spectra are calculated numerically for dimensionless string tensions $G \mu/c^2$ between $10^{-7}$ and $10^{-18}$, and initial loop sizes as a fraction of the Hubble radius $\alpha$ from 0.1 to $10^{-6}$.  The spectra show a low frequency power-law tail, a broad spectral peak due to loops decaying at the present epoch (including frequencies higher than their fundamental mode, and radiation associated with cusps), and a flat (constant energy density) spectrum at high frequencies due to radiation from loops that decayed during the radiation-dominated era.  The string spectrum is distinctive and unlike any other known source.  The peak of the spectrum for light strings appears at high frequencies, significantly affecting predicted signals.  The spectra of the cosmic string backgrounds  are compared with current millisecond pulsar limits and \textit{Laser Interferometer Space Antenna} (LISA) sensitivity curves.  For models with large stable loops ($\alpha=0.1$), current pulsar-timing limits exclude $G \mu/c^2>10^{-9}$, a much tighter limit on string tension than achievable with other techniques, and within the range of current models based on brane inflation.  LISA may detect a background from  strings as light as $G \mu/c^2\approx 10^{-16}$, corresponding to field-theory strings formed at roughly $10^{11}$ GeV.  
  \end{abstract}


\maketitle

       \section{Introduction}

It is well established in gauge field theories that the vacuum expectation value of a field $\left\langle \phi\right\rangle$ can take on non-zero values when the ground state, at the minimum of the zero-temperature potential or ``true vacuum'',  breaks the symmetry of the underlying Lagrangian \cite{vil,hin}.  In the hot early universe, thermal effects lead to an effective potential with a temperature dependence,  such that the vacuum starts in a symmetric ``false'' vacuum and only later cools to its ground state  below a certain critical temperature~\cite{linde,vil,ba,va84,vil85}.  In general however the expansion of the universe occurs too quickly for the system to find its true ground state at all points in space~\cite{ki}.  Depending upon the topology of the manifold of degenerate vacua, topologically stable defects form in this process, such as  domain walls, cosmic strings, and monopoles~\cite{vil,hin,va84,vil85,sak06}.  In particular, any field theory with a broken U(1) symmetry will have classical solutions extended in one dimension, and in cosmology these structures generically form a cosmic network of macroscopic, quasi-stable strings that steadily unravels but survives to the present day, losing energy primarily by gravitational radiation ~\cite{vi81,ho3,vi85,batt,turok,allen01}.  A dual superstring description for this physics is given in terms of one-dimensional branes, such as D-strings and F-strings~\cite{vi05,pol,dv,sas,jones1,pol06}.  

 In this paper we calculate the spectrum of  background gravitational radiation from cosmologies with  strings.  We undertake calculations at much lower string masses than previously, with several motivations:
 
	\begin{enumerate}
\item
Recent advances in millisecond pulsar timing have reached new levels of precision and are providing better limits on low frequency backgrounds; the   calculations presented here provide a precise connection between the background limits and fundamental theories of strings and inflation~\cite{fir05,jenet,jones,tye06}.
   \item
The \emph{Laser Interferometer Space Antenna} (LISA) will provide much more sensitive limits over a broad band around millihertz frequencies.  Calculations have not previously been made for theories in this band of sensitivity and frequency, and are needed since the background spectrum depends significantly on string mass~\cite{lisa,armstrong,ci}.
   \item
Recent studies of string network behavior strongly suggest (though they have not yet proven definitively) a high rate of formation of stable string loops comparable to the size of the cosmic horizon.  This results in a higher net production of gravitational waves since the loops of a given size, forming earlier, have a higher space density. The more intense background means experiments are sensitive to lower string masses~\cite{ring,van2,martins}.
 \item
Extending the observational probes to the light string regime is an important constraint on field theories and superstring cosmology far below the Planck scale.  The current calculation provides a quantitative bridge between the parameters of the fundamental theory (especially, the string tension), and the properties of the observable background~\cite{ki04,sak06}.
\end{enumerate}
The results of our study confirm the estimates made in an earlier exploratory analysis, that also gives simple scaling laws, additional context and background~\cite{ho}.  Here we add a realistic concordance cosmology and detailed numerical integration of various effects.  We estimate that the approximations used to derive the current results are reliable to better than about fifty percent, and are therefore useful for comparison of fundamental theory with real data over a wide range of string parameters. 

Cosmic strings' astrophysical properties are strongly dependent upon two parameters:  the dimensionless string tension $G\mu$ (in Planck units where $c=1$ and $G={m_{pl}}^{-2}$) or $G\mu/c^2$ in SI units, and the interchange probability p.  Our main conclusion is that the current pulsar data~\cite{jenet,det,sti} already place far tighter constraints on string tension than other arguments, such as microwave background anisotropy or gravitational lensing.  Recent observations from WMAP and SDSS \cite{wy} have put the value of $G\mu< 3.5\times10^{-7}$ due to lack of cosmic background anisotropy or structure formation consistent with heavier cosmic strings.  From \cite{wy} it is found that up to a maximum of 7\% (at the 68\% confidence level) of the  microwave anisotropy can be cosmic strings.  In other words, for strings light enough to be consistent with current pulsar limits, there is no observable effect other than their gravitational waves. The limit is already a powerful constraint on superstring and field theory cosmologies.   In the future, LISA will improve this limit by many orders of magnitude.

In general, lower limits on the string tension constrain theories farther below the Planck scale.  In field theories predicting string formation the string tension is related to the energy scale of the theory $\Lambda_s$ through the relation $G\mu\propto\Lambda_s^2/m_{pl}^2$~\cite{vil,ki,ba}.  Our limits put an upper limit on $\Lambda_s$ in Planck masses given by $\Lambda_s<10^{-4.5}$, already in the regime associated with Grand Unification; future sensitivity from LISA will reach $\Lambda_s\approx 10^{-8}$, a range often associated with a Peccei-Quinn scale, inflationary reheating, or supersymmetric B-L breaking scales~\cite{jean}.  In the dual superstring view, some current brane cosmologies predict that the string tension will lie in the range $10^{-6}<G\mu<10^{-11}$~\cite{tye05,fir05,tye06}; our limits are already in the predicted range and LISA's sensitivity will reach beyond their lower bound.

Cosmic strings radiate the bulk of their gravitational radiation from loops~\cite{an,vil,vi85,turok,vi81}.  For a large network of loops we can ignore directionality for the bulk of their radiation which then forms a stochastic background~\cite{allen96}.  Each loop contributes power to the background:
\begin{equation}
\frac{dE}{dt}=\gamma G \mu^2 c,
\end{equation}
where $\gamma$ is generally given to be on the order of 50 to 100 \cite{ho,vil,burden85,ga87,allen94}.  

Gravitational radiation from strings is characterized for our purposes by the dimensionless luminosity of gravitational waves $\gamma$, which depends on the excitation spectrum of typical radiating modes on string loops.  Loop formation is characterized by $\alpha$, the typical size of newly formed loops as a fraction of the Hubble scale $H(t)^{-1}$.  Estimates based on numerical simulations have in the past  suggested very small values, leading to the hypothesis that $\alpha \sim G \mu$ or smaller~\cite{sie2,bennett88,allen90}; those ideas suggest that all but a fraction $\sim G \mu$ or smaller of loops shatter into tiny pieces.  Recent simulations designed to avoid numerical artifacts suggest a radically different picture, that in fact a significant fraction of loops land on stable non-self-intersecting orbits at a fraction $\sim0.1$ of the horizon scale~\cite{van1,van2,ring,martins}.  Our study is oriented towards this view, which leads to a larger density of loops and a more intense background for a given string mass.

For units of the gravitational energy in a stochastic background a convenient measure we adopt is the conventional dimensionless quantity given by
\begin{equation}
\label{eqs:omega}
\Omega_{gw}(f)=\frac{1}{\rho_c}\frac{d\rho_{gw}}{d\ln f},
\end{equation}
where $\rho_{gw}$ is the energy density in the gravitational waves and $\rho_c$ is the critical density.  $\Omega_{gw}(f)$ gives the energy density of gravitational waves per log of the frequency in units of the critical density.  $\Omega_{gw}$ is proportional to the mean square of the strain on spacetime from a stochastic background of gravitational waves.  

There has been much interest in cusps and kinks as sources of gravitational radiation from cosmic strings \cite{da00,da01,da05,sie}.  In this paper an estimate of  the radiation power from cusps and kinks  is included as higher frequency behavior in the loop gravitational wave spectrum,  but the beaming and bursts of gravitational waves is not discussed explicitly.  As estimated in \cite{ho}, in the regime of very light strings, observable bursts are expected to be rare, and harder to detect than the stochastic background.

Finally we would like to state some additional uncertainties associated with the model of strings described below:

\begin{itemize}

\item{ There is the possibility of multiple scales in loop formation: some string theories suggest multiple
stable intersection points and loops connected by these intersections. This may strongly affect the
accuracy of the one scale model, and indeed some of these theories also have significant other modes of decay that  invalidate the  
entire framework used here~\cite{jackson}.}
\item{ The reconnection probability p can be less than one. This may affect loop formation and the gravitational wave background. In general it is thought that this effect always increases the the predicted background although it is not agreed by how much~\cite{dv}.}
\item{ There may be transient behavior of cosmic strings due to the formation process of cosmic quenching.   The model below  
assumes that transient effects have long since settled out by the time our model begins; this seems likely at LISA frequencies which  
correspond to initial loop sizes much larger than the horizon at the end of inflation~\cite{jones1}.}
\end{itemize}
Further numerical string network studies will be needed to resolve these issues.

           \section{Model of String Loop Populations}

  \subsection{Loop Formation}

The behavior of strings on cosmological scales has been thoroughly discussed in the literature~\cite{vil,vil85,hin,ho3,van1,van2,turok1984,allen90,bennett88}.  By the Kibble mechanism strings are created as a random walk with small coherence length, and they evolve following the Nambu-Goto action.  The strings form in large threadlike networks \cite{van1}, and on scales longer than the Hubble length $cH(t)^{-1}$ the network is frozen and stretches with expansion.  The strings move with speed c, and they interact to form loops with a probability p by different mechanisms: one involves two strings, the other a single string forming a loop.  These loops break off from the infinite string network and become the dominant source of gravitational waves.  They oscillate with a fundamental frequency given by $f=2c/L$, where L is the length of the loop.  In general the loops form a discrete set of frequencies with higher modes given by  $f_n=2nc/L$.

The exchange probability p measures the likelihood that two crossing string segments interact and form new connections to one another.  If p=0 then the string segments simply pass through each other; but if p=1 then they can exchange ``partners'', possibly forming a loop.  The value of p depends upon the model~\cite{dv,hash,eto}; it is close to unity in models most commonly discussed but in principle is an independent parameter (and in a broad class of models, the only one) that can modulate the amplitude of the spectrum for a given string tension.  In this paper it is taken to be p=1 and is parameterized as part of the number of loops formed at a given time~\cite{ho,vi81,ho3}.  The number of loops is normalized to the previous results of R.R. Caldwell and B. Allen~\cite{ca}, which is described in detail later in this section. 

  \subsection{Loop Radiation}

For the dynamics of a string the Nambu-Goto action is used:
\begin{equation}
S=-\mu \int \sqrt{-g^{(2)}} d\tau d\sigma,
\end{equation}
where $\tau$ and $\sigma$ are parameterized coordinates on the world sheet of the loop, and $g^{(2)}$ is the determinant of the induced metric on the worldsheet.  The energy-momentum tensor of the string can be given by \cite{vil},
\begin{equation}
T^{\mu\nu}(t,\textbf{x})=\mu \int d\sigma \left(\dot{x}^{\mu}\dot{x}^{\nu}-\frac{dx^{\mu}}{d\sigma} \frac{dx^{\nu}}{d\sigma} \right) \delta^{(3)}(\textbf{x}-\textbf{x}(t,\sigma)),
\end{equation}
where $\tau$ is taken along the time direction.  The radiated power from a loop at a particular frequency, $f_n$, into a solid angle $\Omega$ is,
\begin{equation}
\frac{d\dot{E_n}}{d\Omega}=4\pi G f_n^2 \left(T_{\mu\nu}^*(f_n,\textbf{k}) T^{\mu\nu}(f_n,\textbf{k})-\frac{1}{2} \left|T_{\nu}^{\nu}(f_n,\textbf{k})\right|^2 \right).
\end{equation}
Here $T_{\mu\nu}(f_n,\textbf{k})$ is the Fourier transform of $T_{\mu\nu}(t,\textbf{x})$ \cite{vil}.  Dimensionally the power is given by $G\mu^2$~\cite{turok} with numerical coefficients determined by the frequency mode and the luminosity of the loop.

In general for the power output of a loop at a given mode we can write $\dot{E_n}=P_n G \mu^2 c$~\cite{vil,ca}, where directionality is ignored by effectively inserting the loop into a large network of loops and the $P_n$ are dimensionless power coefficients.  Note that c has been reinserted.  The power coefficients are generally given the form $P_n\propto n^{-4/3}$ \cite{vil} which will create an $f^{-1/3}$ high frequency tail to the power spectrum of a loop.  The total power from a loop is a sum of all the modes:
\begin{subequations}
\begin{eqnarray}
\dot{E}&=&\sum_{n=1}^{\infty}P_n G \mu^2 c,\\
       &=&\gamma G \mu^2 c.
\end{eqnarray}
\end{subequations}
Thus $\gamma$ measures the output of energy and is labeled the ``dimensionless gravitational wave luminosity'' of a loop.  For most calculations we use the value $\gamma=50$~\cite{allen94}.

For the majority of our calculations we assume only the fundamental mode contributes, effectively a delta function for $P(f)$, and $f=2c/L$ is the only emission frequency of each loop.  This assumption is relatively good.  To check this, some calculations   include higher frequency contributions and show that the effect on the background is small overall, with almost no difference at high frequencies.  This is discussed in more detail in Results.  The high frequency additions also represent the contribution from kinks and cusps, which add to the high frequency region with an $f^{-1/3}$ tail from each loop.  This high frequency dependence is the same as that for the high frequency modes, so their contributions should be similar, particularly for light strings.  Light strings contribute at higher frequencies because the loops decay more slowly and more small loops are able to contribute at their fundamental frequency longer throughout cosmic history.

The populations of loops contributing to the background fall naturally into two categories~\cite{ho}: the high redshift ``H'' population of long-decayed loops, and the present-day ``P'' population, where the redshift is of the order unity or less.  The spectrum from the H population, which have long decayed and whose radiation is  highly redshifted, is nearly independent of frequency for an early universe with a $p=\rho/3$ equation of state.  The P loops have a spectrum of sizes up to the Hubble length;  they create a peak in the spectrum  corresponding approximately to the fundamental mode of the  loops whose decay time is the Hubble time today, and a high frequency tail  mainly from higher modes of those loops.  For all of the observable loops (that is, less than tens of light years across), formation occurred during radiation era (that is, when the universe was less than $~10^5$ years old).  As in Hogan \cite{ho} we use scaling to normalize to the high frequency results from R.R. Caldwell and B. Allen \cite{ca}:  $\Omega_{gw}=10^{-8} (G\mu/10^{-9})^{1/2} p^{-1} (\gamma \alpha/5)^{1/2}$, where p=1.  At the high frequency limit of our curves this relation is used to normalize the parameterization for the number of loops created at a given time.

  \subsection{Loop Lengths}

Loop sizes are approximated by the ``one-scale'' model in which the length of created loops scale with the size of the Hubble radius.   The infinite strings create loops with a size that scales with the expansion, creating a distribution with a characteristic size $\alpha/H$.  

The expression for the average length of newly formed loops is $\left<L(t)\right>=\alpha c H(t)^{-1}$, where $\alpha$ measures the length of the loop as a fraction of the Hubble radius.  It is assumed that the newly formed, stably orbiting loops form a range of lengths at a given time, and this range peaks near $\alpha$.  For the numerical calculations, and to facilitate comparison with previous work, we use a delta function for the length.  Thus only one length of loop is created at a given time and this is the average length defined above.  Because of the averaging introduced by the expansion, this introduces only a small error unless the distribution of sizes is large, on the order of several orders of magnitude or more~\cite{ca}, which would no longer be a one scale model.  As a check, models of loop formation with several loop sizes are also analyzed, and it is found that the larger loops tend to dominate the background over the smaller.

The loops start to decay as soon as they are created at time $t_c$, and are described by the equation:
\begin{equation}
\label{eqs:length}
L(t_c,t)=\alpha c H(t_c)^{-1}-\gamma G \mu \frac{t-t_c}{c}.
\end{equation}
(This expression  is not valid as $L\rightarrow0$ but is an adequate approximation.  As the length approaches zero Eqn.~\ref{eqs:length} becomes less accurate, but in this limit the loops are small and contribute only to the high frequency red noise region of the gravitational wave background.  This region is very insensitive to the exact nature of the radiation so Eqn.~\ref{eqs:length} is accurate within the tolerances of the calculations.)  

Loops that form at a time $t_c$ decay and eventually disappear at a time $t_d$.  This occurs when $L(t_c,t_d)=0$ and from Eqn.~\ref{eqs:length}:
\begin{subequations}
\begin{eqnarray}
0&=&\alpha c H(t_c)^{-1}-\gamma G \mu \frac{t_d-t_c}{c},\\
t_d&=&t_c+\frac{\alpha c^2 H(t_c)^{-1}}{\gamma G \mu}.
\end{eqnarray}
\end{subequations}
At some time $t=t_d$ there are no loops created before time $t_c$, so they no longer contribute to the background of gravitational waves.  This is discussed in more detail below. 

  \subsection{Loop Number Density}

For the number density of strings we parameterize the number created within a Hubble volume at a given time by $N_t / \alpha$.  The newly formed loops redshift with the expansion of the horizon by the inverse cube of the scale factor $a(t)^{-3}$:
\begin{eqnarray}
n(t,t')= \frac{N_t}{\alpha} \left(\frac{H(t)}{c}\right)^3 \left(\frac{a(t)}{a(t')}\right)^3.
\end{eqnarray}
$n(t,t')$ is the number density at time $t$ as seen by an observer at time $t'$, and the redshift has been incorporated into the function.  In the Appendix there is a description of the calculation of the scale factor.  Analyzing the high frequency background equation given by C.J. Hogan \cite{ho} from R.R. Caldwell and B. Allen \cite{ca} the curves are normalized by setting $N_t=8.111$.  

At an observation time $t'$ the total number density of loops is found by summing over previous times $t$ excluding loops that have disappeared,
\begin{equation}
\label{eqs:nsum}
n_s(t')=\sum_{t=t_e}^{t'} n(t,t'),
\end{equation}
where $t_e$ is the time before which all the loops have decayed when making an observation at time $t'$.  The earliest time in our sum, $t_e$, is found by solving Eqn.~\ref{eqs:length}:
\begin{eqnarray}
t'=t_e+\frac{\alpha c^2 H(t_e)^{-1}}{\gamma G \mu},
\end{eqnarray}
at a given $t'$.  Thus $t_e(t')$ tells us that at some time $t'$, all of the loops formed before $t_e(t')$ have decayed and are not part of the sum.  For large loops and light strings $t_e$ trends toward earlier in the history of the universe for a given $t'$, allowing more loops to contribute longer over cosmic history.  For the final calculations, which are described in detail in the next section, $t'$ and the number density are summed over the age of the universe.

As the loops decay the energy lost becomes gravitational radiation, which persists in the universe and redshifts with expansion.  This gravitational wave energy is detectable as a stochastic background from the large number of emitting loops.

               \section{Gravitational Radiation from a Network of Loops}

  \subsection{Frequency of Radiation}

The frequency of radiation is determined by the length of the loop, and has normal modes of oscillation denoted by $n$.  The frequency of gravitational radiation from a single loop is given by,
\begin{eqnarray}
f_n(t_c,t)=\frac{2n c}{L(t_c,t)},
\end{eqnarray}
where $t_c$ is the creation time of the loop and $t$ is the time of observation.  There is also redshifting of the frequency with expansion which must be accounted for as this strongly affects the shaped of the background curves.  The strings radiate most strongly at the fundamental frequency so $n=1$ for most of the calculations.  This also facilitates computation times and introduces only small errors in the amplitude of the background, which is shown in Results.  Ultimately the higher modes do not greatly alter the results at the resolution of the computations.

  \subsection{Gravitational Wave Energy}

A network of loops at a given time $t'$ radiates gravitational wave energy per volume at the rate:
\begin{equation}
\label{eqs:rhogw}
\frac{d\rho_{gw}(t')}{dt}=\gamma G \mu^2 c\;n_s(t').
\end{equation}
Recall that $n_s(t')$ is the number density $n(t,t')$ summed over previous times $t$ at a time of observation $t'$; given by Eqn.~\ref{eqs:nsum}.

To find the total gravitational wave energy density at the present time the energy density of gravitational waves must be found at previous times $t'$ and then summed over cosmic history.  Since the energy density is given by Eqn.~\ref{eqs:rhogw} the total density is found by summing over $n_s(t')$ for all times $t'$ while redshifting with the scale factor, $a^{-4}$.  Thus the total gravitational energy per volume radiated by a network of string loops at the current age of the universe is given by:
\begin{subequations}
\label{eqs:densitytime}
\begin{eqnarray}
\rho_{gw}(t_{univ})=\gamma G \mu^2 c \int^{t_{univ}}_{t_0} \frac{a(t')^4}{a(t_{univ})^4} n_s(t') dt',\\
=\gamma G \mu^2 c \int^{t_{univ}}_{t_0} \frac{a(t')^4}{a(t_{univ})^4} \left(\sum^{t'}_{t=t_e} n(t,t')\right)dt'.
\end{eqnarray}
\end{subequations}
Here $t_{univ}$ is the current age of the universe and $t_0$ is the earliest time of loop formation.  The internal sum is over $t$ and the overall integral is over $t'$.  In practice we can make $t_0$ small to observe higher frequency, H loops, earlier during the radiation era.    
 
Eqs.~\ref{eqs:densitytime} do not have explicit frequency dependence, which is needed to calculate the background gravitational wave energy density spectrum $\Omega_{gw}(f)$ determined by Eqn.~\ref{eqs:omega}.  This requires the gravitational energy density be found as a function of the frequency: the number density as a function of time is converted to a function of frequency at the present time.  This is discussed in detail in the Appendix.

For most calculations it is assumed each length of loop radiates at only one frequency, $f=2c/L$.  The number density, as a function of frequency and time $t'$, is redshifted and summed from the initial formation time of loops $t_0$ to the present age of the universe:
\begin{equation}
\rho_{gw}(f)=\gamma G \mu^2 c \int_{t_0}^{t_{univ}} \frac{a(t')^4}{a(t_{univ})^4} n(f\frac{a(t')}{a(t_{univ})},t') dt'.
\end{equation}
Note that the frequency is also redshifted.

We have checked the effects of including higher mode radiation from each loop.  The time rate of change of the gravitational energy density of loops at time $t'$ is found by summing over all frequencies $f_j$ and weighting with the power coefficients $P_j$: 
\begin{equation}
\frac{d\rho_{gw}(f',t')}{dt'}=\sum_{j=1}^{\infty} P_j G\mu^2 c\: n(f_j',t').
\end{equation}
This is then integrated over cosmic time and redshifted to give the current energy density in gravitational waves:
\begin{eqnarray}
  \rho_{gw}(f)&=&G\mu^2 c \int_{t_0}^{t_{univ}}dt' \frac{a(t')^4}{a(t_{univ})^4}
                 \nonumber\\
                 & &\times\left\{\sum_{j=1}^{\infty} P_j\:n(f_j\frac{a(t')}{a(t_{univ})},t')\right\} .
\end{eqnarray}
The power coefficients $P_j$ are functions of the mode and have the form $P_j\propto j^{-4/3}$~\cite{vil,ca}.  This includes the behavior of cusps and kinks, which also contribute an $f^{-1/3}$ tail to the power spectrum.  Depending upon the percentage of power in the fundamental mode, the sum to infinity is not needed.  As an example, if 50\% of the power is in the fundamental mode only the first six modes are allowed.  In general it is found that the higher modes do not significantly effect the values of the background radiation.  Details are given in Results.

From the gravitational radiation density we can find the density spectrum readily by taking the derivative with respect to the log of the frequency and dividing by the critical density:
\begin{subequations}
\begin{eqnarray}
\Omega_{gw}(f)&=&\frac{f}{\rho_c}\frac{d\rho_{gw}(f)}{df},\\
&=&\frac{8 \pi G f}{3 H_o^2} \frac{d\rho_{gw}(f)}{df}.
\end{eqnarray}
\end{subequations}
A convenient measure is to take $\Omega_{gw} h^2$ to eliminate uncertainties in $H_o=H(t_{univ})=100 h\ km\ s^{-1} Mpc^{-1}$ from the final results.

               \section{Results}

  \subsection{General Results}
    
\begin{figure*}
\includegraphics[width=.9\textwidth]{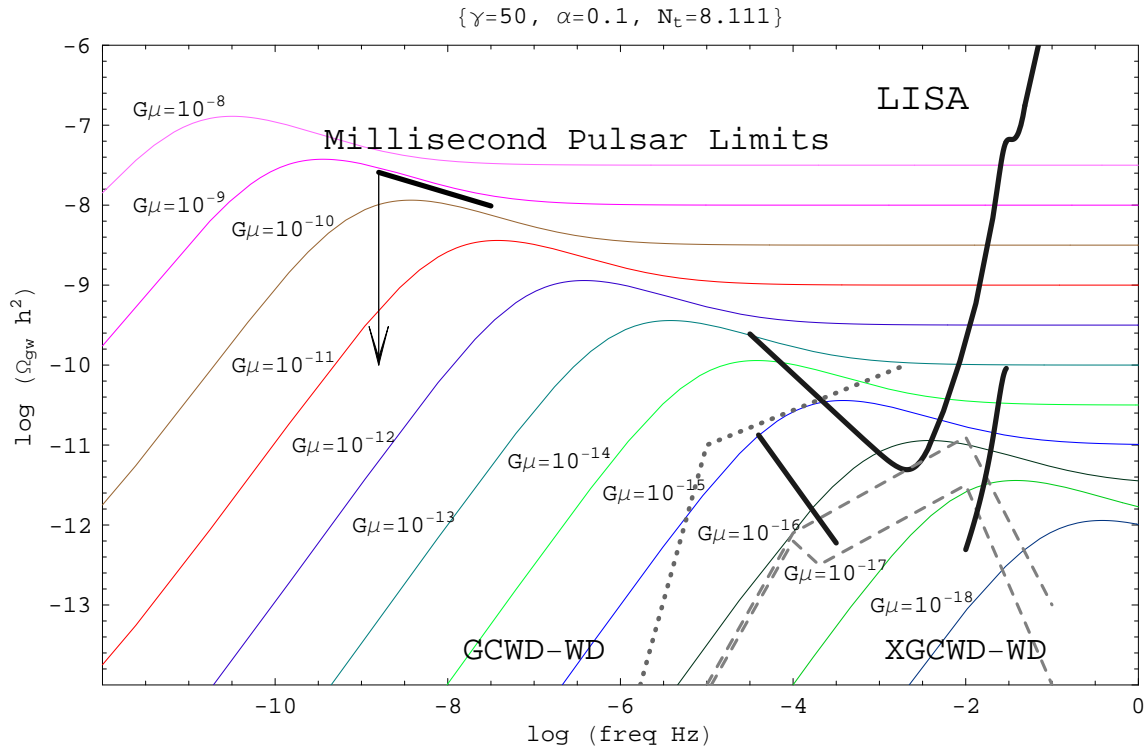}
 \caption{\label{fig1} The gravitational wave energy density per log frequency from cosmic strings is plotted as a function of frequency for various values of $G\mu$, with $\alpha=0.1$ and $\gamma=50$.  Note that current millisecond pulsar limits have excluded string tensions $G\mu>10^{-9}$ and LISA is sensitive to string tensions $\sim10^{-16}$ using the broadband Sagnac technique, shown by the bars just below the main LISA sensitivity curve.  The dotted line is the Galactic white dwarf binary background (GCWD-WD) from A.J. Farmer and E.S. Phinney, and G. Nelemans, et al.,~\cite{farm,nel}.  The dashed lines are the optimistic (top) and pessimistic (bottom) plots for the extra-Galactic white dwarf binary backgrounds (XGCWD-WD)~\cite{farm}.  Note the GCWD-WD eliminates the low frequency Sagnac improvements.  With the binary confusion limits included, the limit on detectability of $G\mu$ is estimated to be $>10^{-16}$.}
\end{figure*}

All of the computations show the expected spectrum for loops:  old and decayed ``H'' loops leave a flat spectrum at high frequencies and matter era ``P'' loops contribute to a broad peak, with smooth transition region connecting them that includes comparable contributions from both.  The frequency of the peak is strongly dependent upon the string tension, with lighter strings leading to a higher frequency peak.  This is a result of the lighter loops decaying at a lower rate, so smaller loops survive for a Hubble time.  This behavior is exhibited in Fig.~\ref{fig1}.  

Fig. 2 shows the peaks of power output per unit volume per log frequency as a function of frequency at different times in cosmic history.  The most recent times contribute to the broad peak, while later times contribute to the flat red noise region.  Although the power output per volume is larger earlier, the summation time is foreshortened so the contributions overlap to form the lower amplitude flat region in the background.

\begin{figure}
  \includegraphics[width=.48\textwidth]{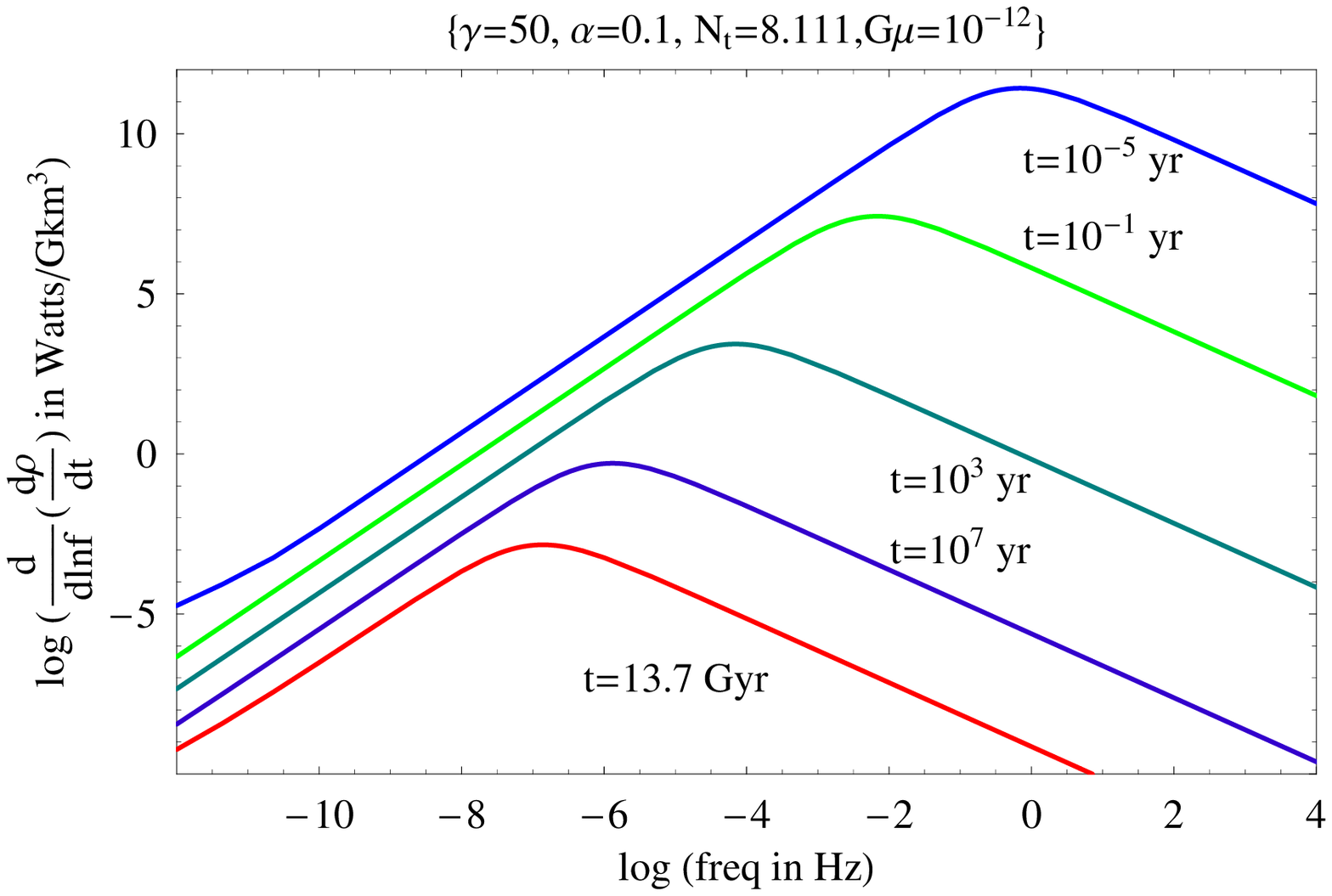}
  \caption{\label{fig2}The power per unit volume of gravitational wave energy per log frequency from cosmic strings is measured at different times in cosmic history as a function of frequency for $G \mu=10^{-12}$, $\alpha=0.1$, and $\gamma=50$.  Note that early in the universe the values were much larger, but the time radiating was short.  The low frequency region is from current loops and the high frequency region is from loops that have decayed. This plot shows the origin of the peak in the spectrum at each epoch; the peaks from earlier epochs have smeared together to give the flat high frequency tail observed today.}
\end{figure}

The differences in these calculations from previous publications can be accounted for by the use of larger loops and lighter strings as predicted in~\cite{ho}.  The change in cosmology, i.e. the inclusion of a cosmological constant, is not a cause of significant variation.  Since the cosmological constant becomes dominant in recent times the increased redshift and change in production rate of loops does not strongly affect the background.

   \subsection{Varying Lengths}

The effect of changing the size of the loops is shown in Fig. 3.  A larger $\alpha$ leads to an overall increase in the amplitude of $\Omega_{gw}$ but a decrease in the amplitude of the peak relative to the flat portion of the spectrum. As expected there is little if no change in the frequency of the peak.  In general the red noise portion of the spectrum scales with $\alpha^{1/2}$ and this is seen in the high frequency limits of the curves in Fig. 3 \cite{ho}.  An increase in $\alpha$ leads to longer strings and an increase in the time that a given string can radiate.  This leads to an overall increase in amplitude without an effect on the frequency dependence of the spectrum peak.

\begin{figure}
 \includegraphics[width=.48\textwidth]{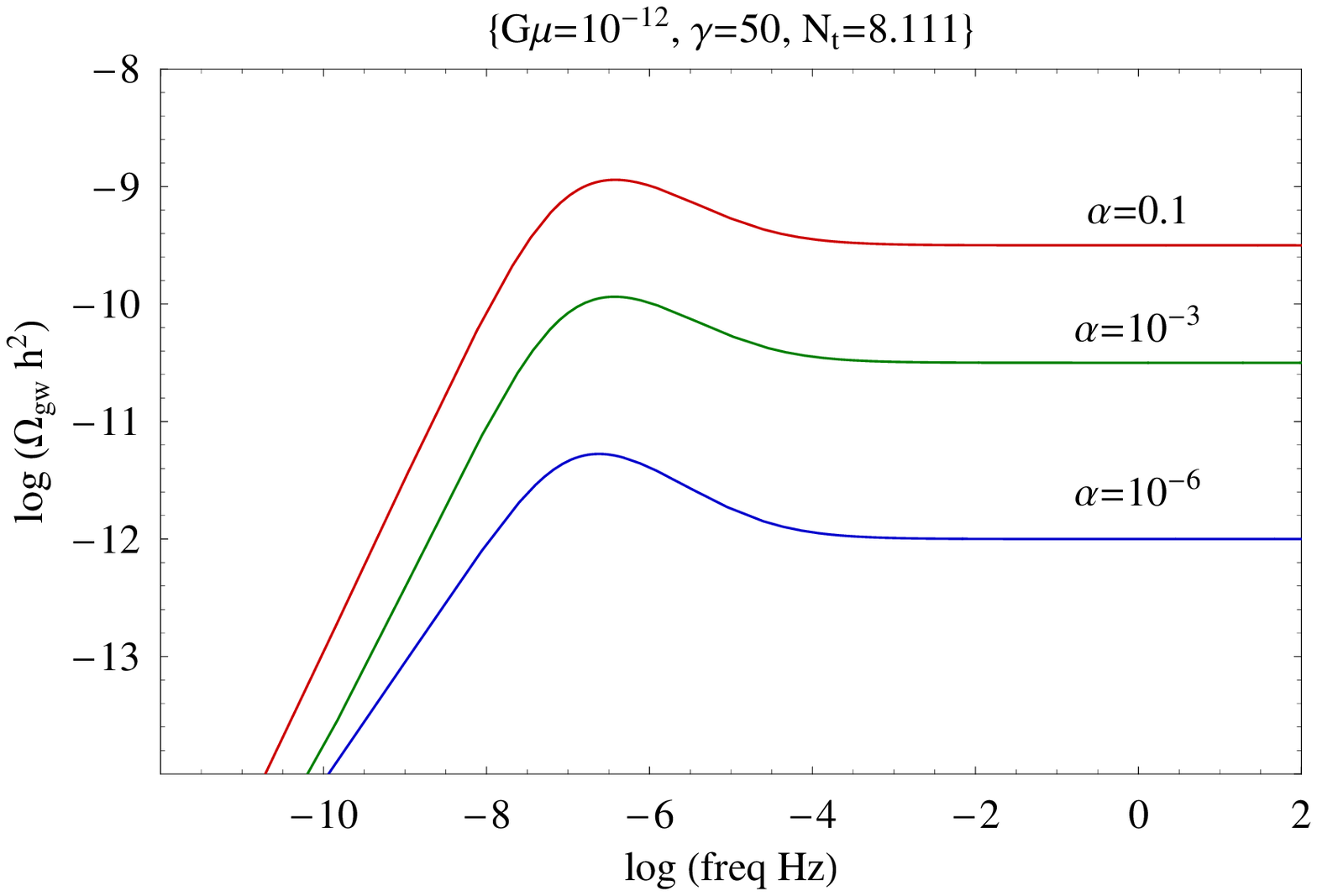}
 \caption{\label{fig3} The gravitational wave energy density per log frequency from cosmic strings is shown with varying $\alpha$ for $G\mu=10^{-12}$ and $\gamma=50$.  Here $\alpha$ is given values of 0.1, $10^{-3}$, and $10^{-6}$ from top to bottom.   Note the overall decrease in magnitude, but a slight increase in the relative height of the peak as $\alpha$ decreases; while the frequency remains unchanged.  The density spectrum scales as $\alpha^{1/2}$, and larger loops dominate the spectrum over small loops.}
\end{figure}

\begin{figure}
 \includegraphics[width=.48\textwidth]{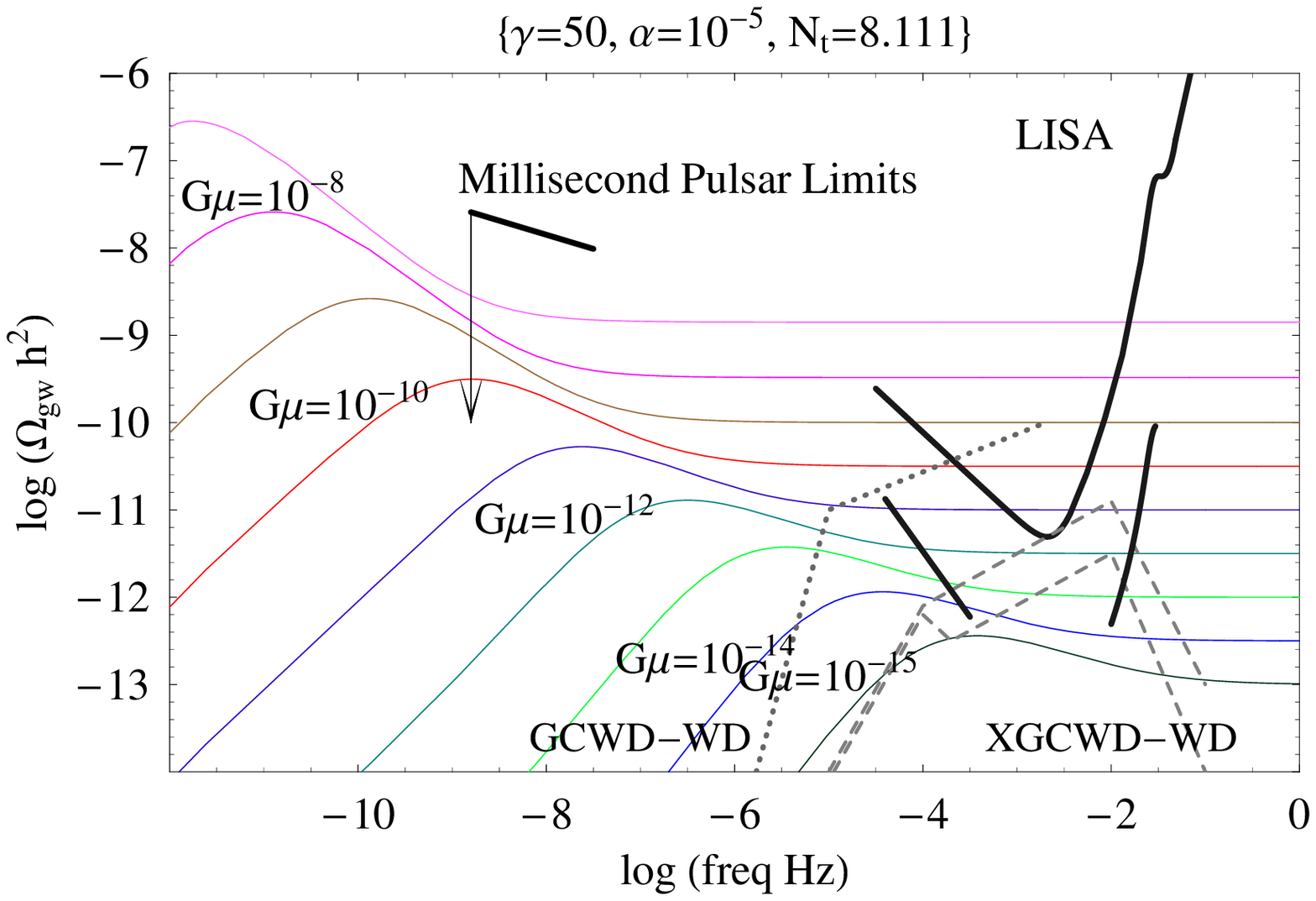}
 \caption{\label{fig4}The gravitational wave energy density per log frequency from cosmic strings for $\alpha=10^{-5}$ and $\gamma=50$.  Smaller $\alpha$ reduces the background for equivalent string tensions, with a LISA sensitivity limit of $G\mu>10^{-14}$ if the Sagnac sensitivity limits are used.  The bars below the main LISA sensitivity curve are the Sagnac improvements to the sensitivity, while the dotted line is the Galactic white dwarf binary background (GCWD-WD) from A.J. Farmer and E.S. Phinney, and G. Nelemans, et al.,~\cite{farm,nel}.  The dashed lines are the optimistic (top) and pessimistic (bottom) plots for the extra-Galactic white dwarf binary backgrounds (XGCWD-WD)~\cite{farm}.  Note the GCWD-WD eliminates the low frequency Sagnac improvements, and increases the minimum detectable $G\mu$ to $>10^{-12}$.  $G\mu$ ranges from the top curve with a value $10^{-7}$ to the bottom curve with $10^{-15}$.} 
 \end{figure}

In Fig.~\ref{fig4} a smaller $\alpha$ is plotted and shows the reduction in the background.  The general shape of the background remains unchanged, except for some small differences in the heavier strings.  For heavier strings the amplitude of the peak tends to increase relative to the red noise portion of the background spectrum, and the frequency of the peak shifts less with $G\mu$.  These differences become more pronounced for heavy and very small strings; and at this limit our results are found to match very closely those computed in~\cite{ca}. 

   \subsection{Luminosity}
\begin{figure}
 \includegraphics[width=.48\textwidth]{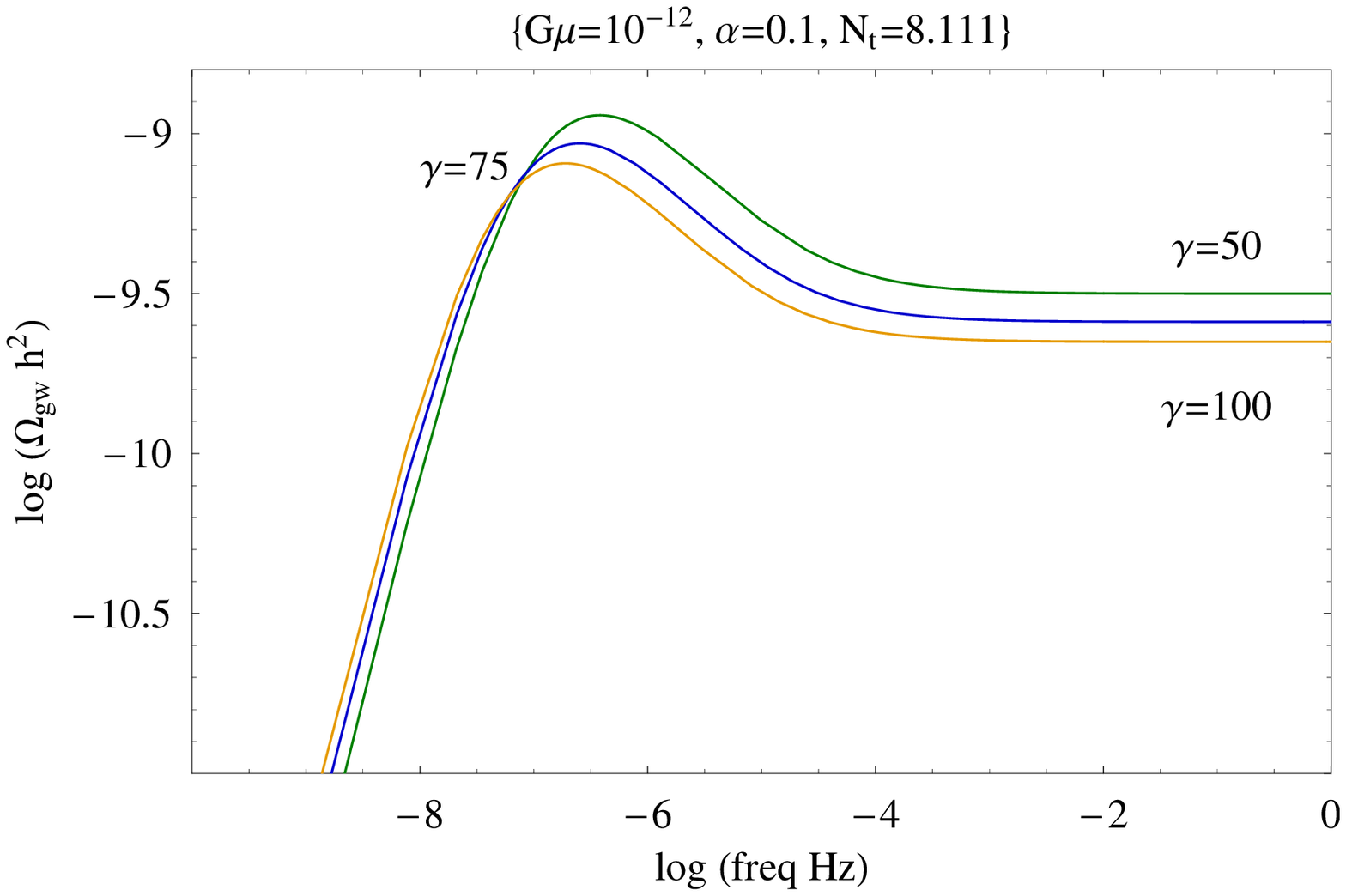}
  \caption{\label{fig5}The energy density in gravitational waves per log frequency from cosmic strings with varying luminosity $\gamma$ for $G\mu=10^{-12}$ and $\alpha=0.1$.  Here the $\gamma$'s are given as follows at high frequencies: 50 for top curve, 75 for the middle curve, and 100 for the bottom curve.  The peak frequency, peak amplitude, and amplitude of flat spectrum all vary with the dimensionless gravitational wave luminosity $\gamma$.}
 \end{figure}

A residual factor of the order of unity arises from uncertainty in the typical radiation losses from loops when all modes are included averaged over an entire population.  Changing the dimensionless luminosity $\gamma$ affects the curves as shown in Fig.~\ref{fig5}.  Larger $\gamma$ decreases the amplitude of the red noise region of the spectrum, but increases the lower frequency region.  Smaller $\gamma$ increases the amplitude of the peak and the flat portion of spectrum, while leading to a decrease in the high frequency region.  This occurs because the loops decay more quickly for larger $\gamma$ and have less time to contribute over the long time periods of cosmology.  In the matter dominated era the higher luminosity allows the loops to contribute more to the high frequency region, although it does not increase the peak amplitude.  Higher luminosity also decreases the frequency of the overall peak by contributing more power in current loops.  The power and change in length depend on the luminosity by $\dot{E}\propto\gamma$ and $\dot{L}\propto-\gamma$.  Note that the gravitational wave power depends on the square of the mass density but the first power of the luminosity, thus different behavior is expected varying $\gamma$ or $\mu$.  An analytic description of the scaling of the amplitude and frequency of $\Omega_{gw}$ with  $\gamma$ is given in~\cite{ho}. 

  \subsection{Multiple Lengths}

\begin{figure}
    \includegraphics[width=.48\textwidth]{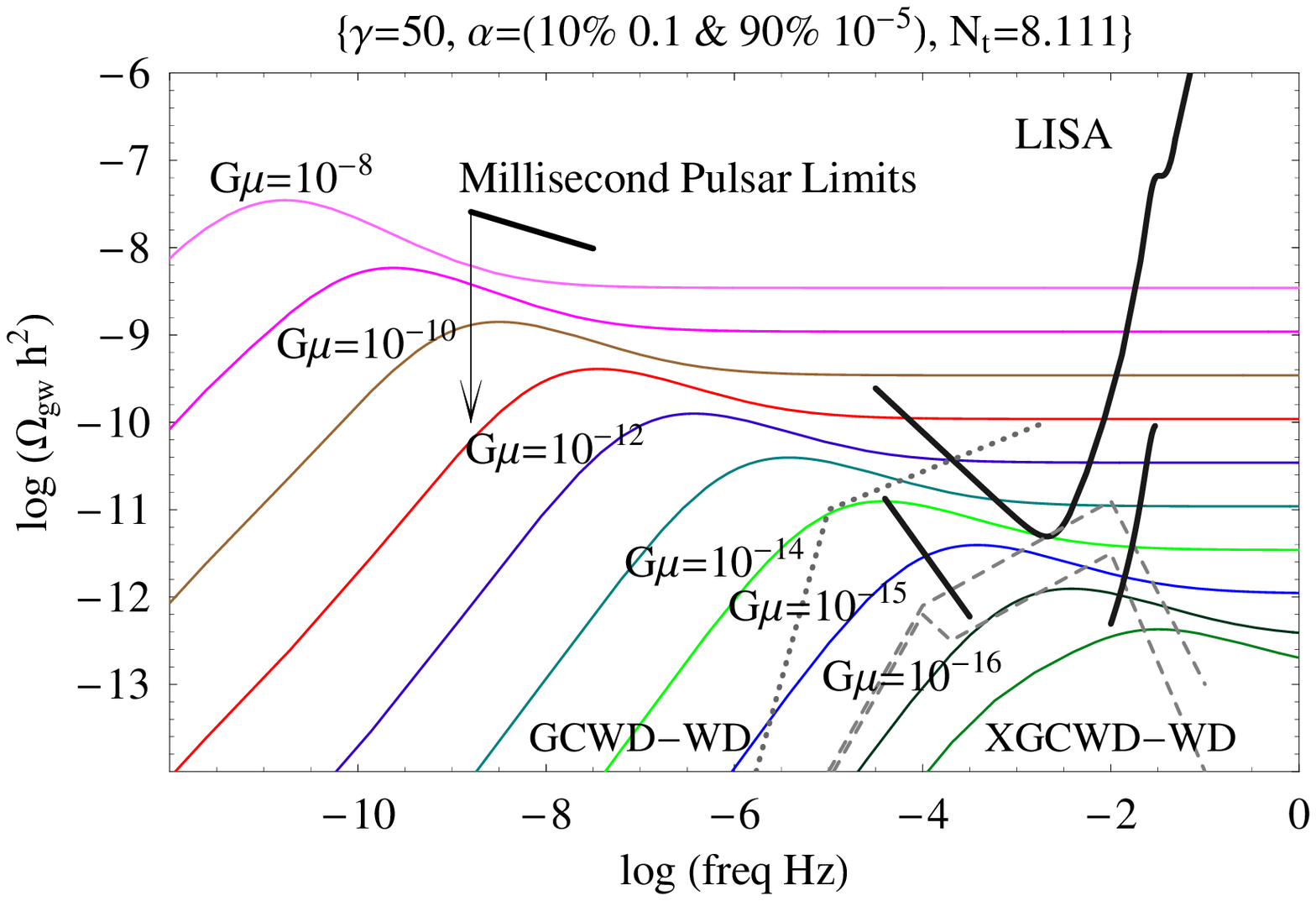}
       \caption{\label{fig6} The gravitational wave energy density per log frequency from cosmic strings for $\gamma$=50, $G\mu=10^{-12}$, and a combination of two lengths ($\alpha$'s).  90\% of the loops shatter into loops of size $\alpha=10^{-5}$ and 10\% remain at the larger size $\alpha=0.1$.  The overall decrease is an order of magnitude compared to just large alpha alone.  From bottom to top the values of $G\mu$ run from $10^{-17}$ to $10^{-8}$. }
    \end{figure}

Some models indicate several characteristic scales for the loop populations, with some combination of several scales.  This implies multiple distributions of lengths, which can be described as a combination of different $\alpha$'s.  Figures~\ref{fig1}, \ref{fig3}, and \ref{fig4} indicate the eventual dominance of the larger loops over the smaller in the final spectrum.  Even though, in this model, the number of loops created goes as $\alpha^{-1}$, their contribution to the background is still much smaller and scales as $\alpha^{1/2}$.  Thus an admixture of loop scales tends to be dominated by the largest stable radiating loops, which is shown in Fig.~\ref{fig6}.

In Fig.~\ref{fig6} we assume 90\% of the loops shatter into smaller sizes $\alpha=10^{-5}$ while 10\% are larger with $\alpha=0.1$.  A ``two scale'' model is used and two relatively narrow distributions are assumed, as in the one scale model.  It is imagined that the inital loops are large and splinter into smaller loops, with the percentage that shatter depedent upon the model.  Overall the amplitude decreases by an order of magnitude below a pure $\alpha=0.1$ amplitude, but it is an order of magnitude above the $\alpha=10^{-5}$ spectra.  Thus the large loops, even though they form a small fraction, contribute more to the background.  Because of this weighting a one-scale model is likely to be suitable if appropriately normalized.  Note that a string simulation requires a dynamic range of scale greater than $\alpha$ to make a good estimate of $\alpha$ and to choose the right normalization.

    \subsection{High Frequency Modes, Kinks, and Cusps}

\begin{figure}
    \includegraphics[width=.48\textwidth]{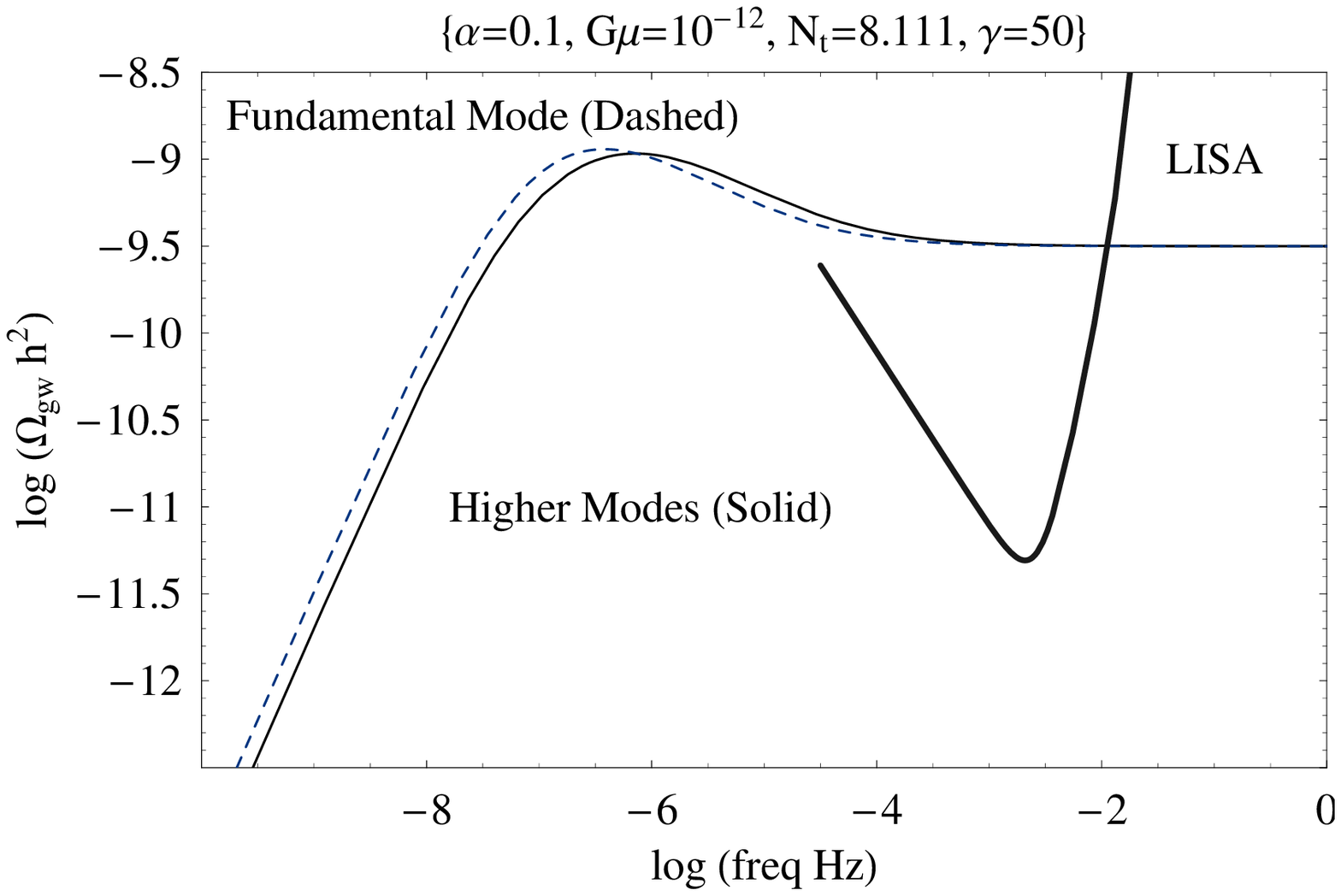}
       \caption{\label{fig7}The gravitational wave energy density per log frequency for $\gamma$=50, $G\mu=10^{-12}$, $\alpha=0.1$; plotted in one case (dashed) with only the fundamental mode, while for another case (solid) with the first six modes.  The dashed curve has all power radiated in the fundamental mode, while the solid curve combines the fundamental with the next five modes: $P_1=25.074,\ P_2=9.95,\ P_3=5.795,\ P_4=3.95,\ P_5=2.93,\ P_6=2.30$.  Note the slight decrease in the peak and its shift to a higher frequency.  The LISA sensitivity is added as a reference.}
    \end{figure}

Calculations are done which include higher mode behavior of the loops; this sends some of the gravitational wave energy into higher frequencies for each loop.  Adding higher modes tends to smooth out, to a larger degree, the curves and slightly lower the peaks on the spectra.  When applied to the power per volume per log frequency as in Fig.~\ref{fig2} we see a less steep high frequency tail at each time and a slight decrease in the amplitude of the peak.  From Fig.~\ref{fig7} a comparison of the fundamental mode versus the fundamental mode with higher modes included shows: a) the peak decreases, b) the peaked portion widens slightly, and c) the red noise amplitude remains relatively unchanged.  

Fig.~\ref{fig7} demonstrates a calculation in which $\sim50\%$ of the power is in the fundamental mode and the power coefficients go as $n^{-4/3}$ in the higher modes.  For the fundamental mode $P_1=25.074$, and the next five modes are $P_2=9.95$, $P_3=5.795$, $P_4=3.95$, $P_5=2.93$, and $P_6=2.30$.  Their sum is $\sum P_n=\gamma=50$.  The amplitude of the peak shifts from $1.202\times10^{-9}$ to $1.122\times10^{-9}$, a 6.7\% decrease.  The frequency of the peak shifts more substantially, from $3.162\times10^{-7}\:Hz$ to $7.080\times10^{-7}\:Hz$, an increase of 124\%.  This increase in frequency of the peak does not affect the limits of detection by LISA because the amplitude shrinks very little.  Note that the high frequency red noise region is essentially unchanged, and for $G\mu=10^{-12}$ this is the region in which LISA is sensitive. 

Plots of spectra for other $G \mu$ with high frequency modes show changes similar to those in Fig.~\ref{fig7}.  The peak amplitude drops slightly and the frequency of the peak is increased.  Overall, the limits from Table~\ref{table1} remain unchanged by the introduction of 50\% of the power in the higher frequency modes.  

Cusps and kinks in the loops are responsible for high frequency ``bursts'' of gravitational wave energy~\cite{da00,da01,da05,sie}.  They are two different manifestations of string behavior: cusps are catastrophes that occur approximately once every oscillation, emitting a directed burst of gravitational radiation; while kinks are small wrinkles which propagate on the loops, emitting a higher frequency of directed gravitational energy.  The light strings make it less likely to detect these random bursts; in particular they are more difficult to pick out from their own confusion background~\cite{ho}.  Since cusps, kinks, and higher modes add an $f^{-1/3}$ tail to the power spectrum of each loop, it is reasonable to expect that they will have a similar effect on $\Omega_{gw}$.  The general behavior of high frequency contributions is represented in Fig.~\ref{fig7}.

                               \section{Millisecond Pulsar Limits}

    \subsection{Limits on Background}

Millisecond pulsars have been used as sources to indirectly measure a gravitational radiation background for a number of years \cite{det,sti,kaspi,lommen,thors,mch}.  Limits on the strain, and correspondingly the energy density spectrum $\Omega_{gw}$, have been estimated recently at three frequencies: $1/20\:yr^{-1}$, $1/8\:yr^{-1}$, and $1\:yr^{-1}$~\cite{jenet}.  The limits on the strain vary with the frequency dependence assumed for the source of gravitational radiation at the  corresponding frequencies:
\begin{equation}
\label{eqs:Aheqn}
h_c(f)=A \left(\frac{f}{yr^{-1}}\right)^{\beta},
\end{equation}
where $\beta$ is the power of the frequency dependence of the source and is denoted as $\alpha$ in F.A. Jenet, et al.,\cite{jenet}.  A table of values is provided in that source for various cosmological sources of stochastic gravitational radiation, with $\beta=-7/6$ used for cosmic strings.  From~\cite{jenet}, the cosmic string limits are:
\begin{subequations}
\label{eqs:limits}
\begin{eqnarray}
\Omega_{gw}(1\ yr^{-1}) h^2    &\leq& 9.6\times10^{-9},\\
\Omega_{gw}(1/8\ yr^{-1}) h^2  &\leq& 1.9\times10^{-8},\\
\Omega_{gw}(1/20\ yr^{-1}) h^2 &\leq& 2.6\times10^{-8}.
\end{eqnarray}
\end{subequations}

Results of our calculations show that for heavier strings at the measured frequencies $\beta$ is approximately -1, but for lighter strings it varies from this value.  The variation of $\beta$ is summarized in Table~\ref{table2}, which has been calculated at $f=1/20\:yr^{-1}$ and $f=1\:yr^{-1}$.  This change in $\beta$ will result in limits differing slightly only at low frequencies from those in Eqs.~\ref{eqs:limits} \cite{jenet}.

In Figures~\ref{fig1} and \ref{fig4} the limits for millisecond pulsars \cite{jenet} are shown, with allowed values of $\Omega_{gw}$ indicated by the direction of the arrow.  For large strings $\alpha=0.1$, the maximum allowed $G \mu$ is about $10^{-9}$, Fig.~\ref{fig1}.  Decreasing $\alpha$ decreases the amplitude of the background allowing for heavier strings within the pulsar measurement limits.  A more complete list is displayed in Table~\ref{table1}.

  \subsection{Dependence of Characteristic Strain on Frequency}

From the curves of Fig.~\ref{fig1} one can find the frequency dependence of the characteristic strain as given by $h_c(f)=A(f/yr^{-1})^{\beta}$.  Again, $\beta$ is the power law dependence of $h_c$ on frequency, and is generally written as $\alpha$ with a numerical value of -7/6 \cite{jenet,mag,mag00}.  In this paper $\beta$ is used to prevent confusion with the length of loops.  If $\Omega_{gw}$ is known, then using Eqn.~\ref{eqs:charstrain} we find:
\begin{equation}
\Omega_{gw}(f)=\frac{2 \pi^2 A^2}{3 H_o^2} \frac{f^{2(1+\beta)}}{(3.17\times10^{-8}\:Hz)^{2 \beta}}.
\end{equation}
With this equality $\beta$ is found by computing $d\ln(\Omega_{gw})/d\ln f$ at $f=1\:yr^{-1}=3.17\times10^{-8}Hz$ and setting this equal to $2(1+\beta)$.  The results are listed in Table~\ref{table2}.

\begin{table}
 \caption{\label{table2}$\beta$ for given $G\mu$ at $f=1/20\:yr^{-1}$ and $f=1\ yr^{-1}$, $\alpha=0.1$, $\gamma=50$, where $h_c\propto f^\beta$.}
 \begin{ruledtabular}
 
\begin{tabular}{c c c}
     & $\beta$ & $\beta$ \\
 $G\mu$ & $f=1/20\:yr^{-1}$ & $f=1\:yr^{-1}$ \\
\hline
$10^{-6}$& -1.00306 & -1.00031 \\

$10^{-8}$& -1.05314 & -1.00894 \\

$10^{-9}$& -1.05985 & -1.03853 \\

$10^{-10}$& -0.931749 & -1.06593 \\

$10^{-11}$& -0.735778 & -0.987882 \\

$10^{-12}$& -0.682874 & -0.778679 \\

$10^{-13}$& -0.676666 & -0.688228 \\

$10^{-14}$& -0.676034 & -0.676054 \\

$10^{-15}$& -0.675971 & -0.674794 \\

$10^{-16}$& -0.675965 & -0.674668 \\

 \end{tabular}
 \end{ruledtabular}
 \end{table}

From Fig.~\ref{fig1} the change in slope with $G\mu$ is evident at the measured frequency $f_o=1\ yr^{-1}$.  For large values of $G\mu$, $f_o$ is in the flat red noise section of the spectrum so the slope of $\Omega_{gw}$ goes to zero, and $\beta$ approaches -1.  As the string tension is reduced, $\beta$ is measured first at the high frequency tail so the slope is negative, and thus $\beta$ is more negative.  Once over the peak on the high frequency side, the slope becomes positive, thus $\beta$ must be greater than -1.  The characteristic strain power law dependence $\beta$ then approaches a value of -0.67.  

This dependence on $G\mu$ changes the value of the constant $A$ in Eqn.~\ref{eqs:Aheqn} and thus the limits on $\Omega_{gw}$ \cite{jenet}.  Higher mode dependence only modestly changes the values of $\beta$ since it approaches the limits -1 and -2/3 for high and low frequencies respectively.  The frequency at which it takes on these values depends upon $G \mu$ as seen in Fig.~\ref{fig1} 

                                  \section{Future LISA Sensitivity}

   \subsection{Physical Principles}

The pending launch of LISA will give  new opportunities to observe gravitational waves in general, and a stochastic background in particular~\cite{ci,allen96,cornish,lisa}.  There are a number of theoretical sources for this background, including cosmic strings \cite{mag,kosow,ferrari99,postov,ho4}.  LISA sensitivity has been modeled and methods for improving sensitivity by calibrating noise and integrating over a broad band have been proposed \cite{shane,hoben,corn}.  The cosmic string background density spectrum calculated here is compared to the potential for detection by LISA. 

Detectors such as LISA measure the strain on spacetime, so a relation between the background spectrum and strain is needed.  The rms strain is denoted by $h_{rms}(f)$ or $\bar{h}(f)$, and for an isotropic, unpolarized, stationary background~\cite{jenet,hoben,mag00,allen96}:
\begin{equation}
\Omega_{gw}(f)=\frac{4 \pi^2}{3 H_o^2} f^3 h^2_{rms} (f).
\end{equation}
Another measure is the characteristic strain spectrum $h_c(f)^2=2fh^2_{rms}(f)$, when substituted gives:
\begin{equation}
\label{eqs:charstrain}
\Omega_{gw}(f)=\frac{2 \pi^2}{3 H_o^2} f^2 h_c^2(f).
\end{equation}
In the literature, $S_h(f)=h_{rms}(f)^2$ is often used and referred to as the spectral density \cite{mag00}.  These relations are important as they relate the energy density spectrum $\Omega_{gw}(f)$, with the strain that LISA will detect.  This limit on the detectable strain is shown as a curve on our plots and is taken from~\cite{shane}.  

In this paper the LISA sensitivity curve is calculated with the following parameters: SNR=1.0, arm length=5$\times10^9\:m$, optics diameter=$0.3\:m$, laser wavelength=$1064\:nm$, laser power=$1.0\:W$, optical train efficiency=0.3, acceleration noise=$3\times10^{-15}\:m/(s^2\sqrt{Hz})$, position noise=$2\times10^{-11}\:m/\sqrt{Hz}$.

Extracting the background signal by combining signals over a broad band and an extended time can increase the sensitivity of the detector.  One method known as the ``Sagnac technique'' is shown on plots of the background \cite{hoben,corn}, for example see Fig.~\ref{fig1}.

   \subsection{Limits for LISA Sensitivity}

For both Figures~\ref{fig1} and \ref{fig4} there are limits on the detectability of the background due to strings of various mass densities by LISA from~\cite{shane}.  From Fig.~\ref{fig1} with $\alpha=0.1$ the minimum value of the string tension measurable is $\sim10^{-16}$.  Fig. 4, which plots a smaller $\alpha=10^{-5}$, the minimum is $G\mu\sim10^{-12}$.  These values and others are listed in Table~\ref{table1}.  Sagnac observables increase our available spectrum and change the detectable string tensions as indicated \cite{hoben}. 

Shown in Table~\ref{table1} are the limits on string tensions for string sizes $\alpha$ of 0.1 to $10^{-6}$, along with the corresponding minimum tension that LISA can detect.  Galactic and extra-Galactic binaries confusion limit is taken into account for this table.  For $\alpha<10^{-2}$ the maximum string tension will be greater than $10^{-8}$, although these values are excluded by WMAP and SDSS~\cite{wy}.  It is clear from the table and figures that decreasing $\alpha$ increases the maximum string tension allowed as well as increasing the minimum string tension detectable by LISA.  

 \begin{table}
 \caption{\label{table1}Millisecond Pulsar Limits and LISA sensitivity for $\gamma=50$.  WMAP and SDSS \cite{wy} have excluded $G\mu\geq10^{-7}$.}
 \begin{ruledtabular}
 \begin{tabular}{c c c}
  
            & Millisecond Pulsar  &  Minimum LISA \\
  $\alpha$  & Limits, $G\mu<$   &   Sensitivity, $G\mu\geq$\\   
\hline
0.1        & $10^{-9}$ & $10^{-16}$ \\

$10^{-2}$  & $10^{-8}$  & $10^{-15}$\\

$10^{-3}$  & -          & $10^{-14}$\\

$10^{-4}$  & -          & $10^{-13}$\\

$10^{-5}$  & -          & $10^{-12}$\\

$10^{-6}$  & -          & $10^{-11}$\\

 \end{tabular}
 \end{ruledtabular}
 \end{table}

   \subsection{Confusion Noise from Binary Systems}

Galactic white dwarf binaries add a background ``confusion'' that shrouds portions of LISA's spectrum.  This background has been studied and predictions given for its spectrum~\cite{nel}.  There are also extra-Galactic binary sources which similarly create confusion noise~\cite{farm}.  These confusion limits are shown on plots of the gravitational radiation spectra, and they affect the minimum values of string tension that are detectable.

                                   \section{Conclusions}

Broadly speaking, the results here confirm with greater precision and reliability the estimates given in \cite{ho}.  Our survey of spectra provides a set of observational targets for current and future projects able to detect stochastic gravitational wave backgrounds.  For light strings, the string tension significantly affects the  spectrum of gravitational radiation from cosmic string loops. In addition to a reduction in the overall radiation density,   the most significant change is  the shift of the peak of the spectra to higher frequencies for light strings. Near the current pulsar limits, the peak happens to lie close to the frequency range probed by the pulsars, so spectra as computed here are needed as an input prior for establishing the value of the limits themselves.  For much lighter strings,  the peak is at too high a frequency for strings  to be detectable by pulsar timing as the spectrum falls below other confusion backgrounds.  Lighter strings will be detectable by LISA.   For the lightest detectable strings the peak happens to lie in the LISA band and the detailed spectrum must again be included in mounting observational tests.  The spectrum from strings is quite distinctive and not similar to any other known source.  

A high probability of forming loops with stable, non-self-intersecting orbits, as suggested by recent simulations, leads to larger string loops at formation which in turn  give an increased output of gravitational wave energy and improved possibilities for detection for very light strings.  Recent simulations have shown stable radiating loops form at a significant fraction of the horizon, of order $\alpha=0.1$;  for loops of this size our calculations show that  the maximum string tension allowed by current millisecond pulsar measurements is $G\mu<10^{-9}$, and  the minimum value detectable by LISA above estimated  confusion noise is $G\mu\approx10^{-16}$.  In field theories,  the string tension  is typically related to the scale by $G\mu\propto\Lambda_s^2/m_{pl}^2$, so  the maximum detectable scale in Planck masses currently allowed by millisecond pulsars is $\Lambda<10^{-4.5}$, or around the Grand Unification scale; with LISA, the limit will be about  
 $\Lambda_s\sim10^{-8}$ or $10^{11}\:GeV$.   These results suggest that gravitational wave backgrounds are already a uniquely deep probe into the phenomenology of Grand  Unification and string theory cosmology, and will become more powerful in the future.  The most important  step in establishing these arguments for an unambiguous limit on fundamental theories will be a reliable quantitative calculation of $\alpha$.

                    \appendix*
               \section{Numerical Calculations}

   \subsection{General}

For computations the relevant differential and integral equations are solved numerically.  Here is a list of the equations that are used, most of which have been discussed in the sections above.  In addition there are several more that are expanded upon in this section.  Below note that $n(t,t')$ is the number density with time $t$ at an observation time $t'$, while $n(f',t')$ is the number density as a function of frequency and time $t'$.  The time $t$ is always earlier than the time $t'$, which is the time of observation of the quantity in question.
\begin{eqnarray}
\dot{a}&=&\sqrt{\frac{8 \pi G}{3} \rho a^2},\\
\rho&=&\rho_{rad}+\rho_{\nu}+\rho_{matter}+\rho_\Lambda 
              \nonumber\\
     & &{}+\rho_\infty+\rho_L+\rho_{gw},\\
L(t_c,t)&=&\alpha c H(t_c)^{-1}-G \mu \frac{t-t_c}{c},\\
f(t,t')&=&\frac{2c}{L(t,t')},\\
n(t,t')&=&\frac{N_t}{\alpha} \left(\frac{H(t)}{c}\right)^3 \left(\frac{a(t)}{a(t')}\right)^3,\\
\rho_{gw}(f)&=&\gamma G \mu^2 c \int_{t_0}^{t_{univ}} \frac{a(t')^4}{a(t_{univ})^4} 
            \nonumber\\
              & &\times\:n(f\frac{a(t')}{a(t_{univ})},t') dt'.
\end{eqnarray}
Recent observations allow us to neglect loops $\rho_L$, infinite strings $\rho_{\infty}$, and gravitational waves $\rho_{gw}$ in the Friedmann equation, so these terms are removed \cite{pdg}. 

     \subsection{Scale Factor}

Calculations of the scale factor are done using the standard cosmological model with a homogeneous isotropic spacetime defined by the Friedmann-Robertson-Walker metric, solved using Einstein's equations~\cite{ba,dodson,kolb}.  As an addition the effects of a cosmological constant on the cosmic string gravitational wave background are computed.  The energy densities used to describe the dynamics of the scale factor are radiation, neutrinos, matter, and a cosmological constant.  The equation of motion for the scale factor is determined by the Friedmann equation.
\begin{equation}
\label{friedmann}
\left(\frac{\dot{a}}{a}\right)^2=\frac{8\pi G}{3}(\rho_{rel}+\rho_m)+\frac{\Lambda}{3},
\end{equation}
where the $\rho$'s are functions of the scale factor. We then integrate the above equation to find $a(t)$.

The relativistic and matter densities follow the relations:
\begin{subequations}
\begin{eqnarray}
\rho_{rel}=\frac{(\rho_{rel})_o a_o^4}{a(t)^4},\\
\rho_{m}=\frac{(\rho_{m})_o a_o^3}{a(t)^3}.
\end{eqnarray}
\end{subequations}
The gravitational radiation density also goes as $a(t)^{-4}$. 

In order to compute the scale factor the Friedmann equation is transformed into an integral equation and solved numerically.   The main source of error is the relativistic treatment of neutrinos for all time, but this will incur a very small error in more recent times.  Given that neutrinos have mass, it is possible several species will be non-relativistic at the present time.  This is a minimal consideration as the matter and dark energy contributions are larger by at least five orders of magnitude when the neutrinos are non-relativistic.  The following parameters are taken from the Particle Data Group, 2006~\cite{pdg}:
\begin{subequations}
\begin{eqnarray}
\Omega_{\Lambda}&=&0.76,\\
\Omega_{matter}&=&0.24,\\
  \Omega_{rad}&=&4.6\times10^{-5},\\
 \Omega_{\nu}&=&3.15\times10^{-5},\\
           h&=&0.73,\\
      \rho_c&=&\frac{3 H_o^2}{8 \pi G},
\end{eqnarray}
\end{subequations}
where $\Omega_j=\rho_j /\rho_c$, and the constant $h$ is given by $H_o=100h\ km\ s^{-1}Mpc^{-1}$.

    \subsection{Gravitational Radiation from Loops}

At a given time $t'$, there is an earlier time $t$ before which all of the loops have decayed, which we denote $t_e$.  To find this value at any time $t'$, Eqn.~\ref{eqs:length} is set to zero giving $L(t,t')=0$, and  $t=t_e$ is calculated numerically.  Note that $t_e$ is a function of $t'$.  For very early times $t_e(t')$ can be lower than the initial time in our sum over cosmic history $t_0$.  In this case $t_0$ was used instead of the value for $t_e(t')$. 

For many of the calculations it is assumed each loop radiates at only one frequency, $f=2c/L$.  Then a numerical list of values is computed for $t=t_e$ to $t=t'$, $\{L(t,t'),n(t,t')\}$, and an interpolation function is created from the list.   This is the number density as a function of length and time $t'$, $n(L,t')$.  Next a list of frequency and number density is created, $\{2c/L,n(L,t')\}$, with lengths running from $L(t_e,t')$ to $L(t',t')$.  An interpolation function is made from this list, and this results in the number density as a function of time $t'$, $n(f',t')$.  But the frequencies must be redshifted to the current time $t_{univ}$ so the function is $n(fa(t')/a(t_{univ}),t')$.

Then we must integrate over all $t'$ from the earliest time of loop formation $t_0$ to the present age of the universe $t_{univ}$.  This integral,
\begin{equation}
\rho_{gw}(f)=\gamma G \mu^2 c \int_{t_0}^{t_{univ}} \frac{a(t')^4}{a(t_{univ})^4}\ n(f\frac{a(t')}{a(t_{univ})},t') dt',
\end{equation}
is transformed into a sum,
\begin{equation}
\rho_{gw}(f)=\gamma G \mu^2 c \sum_{t'=t_0}^{t_{univ}} \frac{a(t')^4}{a(t_{univ})^4}\ n(f\frac{a(t')}{a(t_{univ})},t') \Delta t'.
\end{equation}
Suitable $\Delta t'$'s are chosen to ensure both computability and to minimize errors.  In general we solve all equations in logarithmic bins so $\Delta t'$ will vary in accordance.  Using small values of $\Delta t$ to facilitate computation speed will result in relatively small error; choosing $\Delta t=0.1$ in log scale gives a 10\% smaller value for $\Omega_{gw}$ compared to $\Delta t=0.001$ or $\Delta t=0.01$ in log scale.  Values smaller than $\Delta t=0.01$ give little or no change to the background.  After the sum the result is the gravitational wave density as a function of frequency at the present time.

To include higher modes the power coefficients $P_j$ must be included in the calculation.  The subscript $j$ is used instead of $n$ here to prevent confusion with the number density, $n(f',t')$.  At each time $t'$ the number density, and thus the time rate of change of the energy density, of loops is calculated exactly as above, but more frequencies are included.  Each number density at frequency $f_j$ is summed and weighted with $P_j$, and this number density is then integrated over cosmic history,
\begin{eqnarray}
  \rho_{gw}(f)&=&G\mu^2 c \int_{t_0}^{t_{univ}}dt' \frac{a(t')^4}{a(t_{univ})^4}
                 \nonumber\\
                 & &\times\left\{\sum_{j=1}^{\infty} P_j\: n(f_j\frac{a(t')}{a(t_{univ})},t')\right\} .
\end{eqnarray}
The overall integral is then transformed into a sum as before.  The sum of the power coefficients is $\sum P_j=\gamma$, where the $P_j\propto j^{-4/3}$.

The final result is to calculate $\Omega_{gw}(f)$ at the present time.  This is done by finding the current energy density of gravitational waves as a function of frequency, and then taking the derivative with respect to f:
\begin{equation}
\Omega_{gw}(f)=\frac{f}{\rho_c}\frac{d\rho_{gw}}{df}.
\end{equation}
This is then plotted as $\log(\Omega_{gw}(f)\ h^2)$ vs. $\log(f/Hz)$, as seen in Figures 1, $3-7$.  This quantity is, again, the density of stochastic  gravitational wave energy per log of frequency in units of the critical density and is related to the strain on space by the gravity waves.

       \begin{acknowledgments}
The authors wish to thank A. Nelson, L. Yaffe, and M. Strassler for helpful discussions concerning field theory and string theory; and F.A. Jenet and D. Backer for helpful discussions on pulsar timing limits.\end{acknowledgments}


\end{document}